\documentclass[aps,prl,superscriptaddress,amsmath,amssymb,showpacs,twocolumn,letterpaper]{revtex4-1}

\usepackage{graphicx}
\usepackage{dcolumn}
\usepackage{bm}

\begin{document}

\title{Direct Measurement of Auger Electrons Emitted from a Semiconductor Light-Emitting Diode under Electrical Injection: Identification of the Dominant Mechanism for Efficiency Droop}

\author{Justin~Iveland}
\affiliation{Materials Department, University of California, Santa Barbara, CA 93106  USA}

\author{Lucio~Martinelli}
\affiliation{Laboratoire de Physique de la Mati\`ere Condens\'ee, CNRS-Ecole Polytechnique, 91128 Palaiseau Cedex, France}

\author{Jacques~Peretti}
\affiliation{Laboratoire de Physique de la Mati\`ere Condens\'ee, CNRS-Ecole Polytechnique, 91128 Palaiseau Cedex, France}

\author{James~S.~Speck}
\affiliation{Materials Department, University of California, Santa Barbara, CA 93106  USA}

\author{Claude~Weisbuch}
\email{weisbuch@engineering.ucsb.edu}
\affiliation{Materials Department, University of California, Santa Barbara, CA 93106  USA}
\affiliation{Laboratoire de Physique de la Mati\`ere Condens\'ee, CNRS-Ecole Polytechnique, 91128 Palaiseau Cedex, France}

\date{December 25, 2012}

\begin{abstract}
We report on the unambiguous detection of Auger electrons by electron emission spectroscopy from a cesiated InGaN/GaN light emitting diode (LED) under electrical injection. Electron emission spectra were measured as a function of the current injected in the device.  The appearance of high energy electron peaks simultaneously with an observed drop in electroluminescence efficiency shows that hot carriers are being generated in the active region (InGaN quantum wells) by an Auger process.  A linear correlation was measured between the high energy emitted electron current and the ``droop current'' - the missing component of the injected current for light emission.  We conclude that the droop phenomenon in GaN LED originates from the excitation of Auger processes.
\end{abstract}


\maketitle

The Auger carrier recombination mechanism is universal in semiconductors and plays a major role in limiting the performance of devices such as long wavelength telecommunications lasers \cite{*[{See e.g.}]r1} or, with a more limited role, solar cells under high excitation \cite{r2}. The most prevalent mechanism is a three-particle interaction, two electrons and one hole, or two holes and one electron, in which an electron-hole pair recombines releasing its energy nonradiatively by promoting the remaining particle into a higher energy state. 

\begin{figure*}[bt]
\includegraphics[width=16cm]{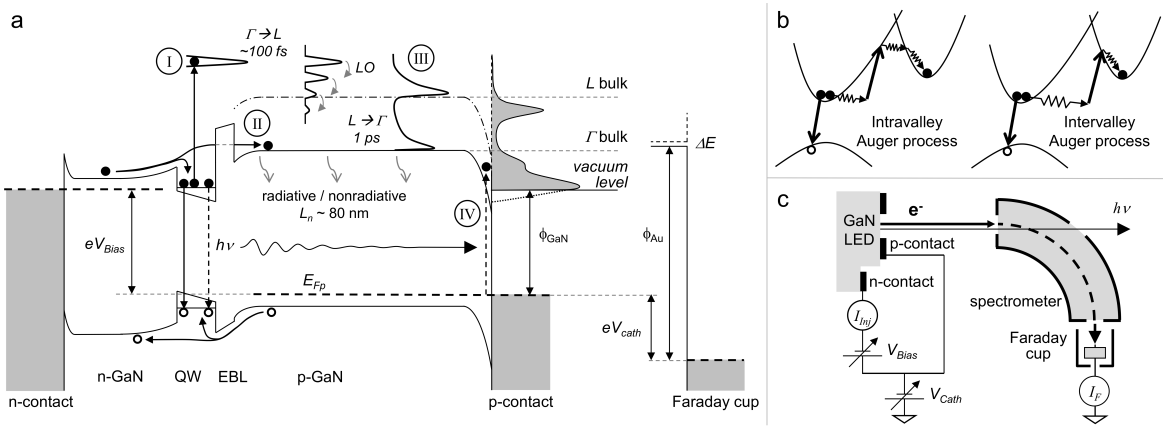}
\caption{\label{fig:1} (a) Principle of band structure and energy levels for a biased LED structure emitting electrons in vacuum. (b) Schematics of hot electron generation into $L$ valley by an eeh Auger process; left: the Auger electron is created in the $\Gamma$ band (intravalley process) and transferred to the $L$ valley; right: the Auger electron is created in the $L$ valley (intervalley process). (c) Schematics of the energy analysis setup for measurement of electrons emitted into vacuum.}
\end{figure*}

In GaN light emitting diodes (LEDs), Auger recombination is also invoked as a possible origin of the so called efficiency droop phenomenon, i.e., drop in light emission quantum efficiency at high carrier injection \cite{r3}. However, the interpretation of the droop phenomenon remains highly controversial, in spite of its importance and many studies. Many proposed mechanisms rely on the enhanced nonradiative (NR) recombination at point defects in either quantum barriers or surrounding majority carrier regions when carriers are no longer localized in the high radiative recombination efficiency regions of quantum wells (QWs): carrier overflow from the QWs into regions of efficient NR recombination \cite{r4,r5} in particular due to saturation of localized states in the QW’s \cite{r6}; loss of current injection efficiency \cite{r5}; density-activated defect recombination \cite{r7}; insufficient hole injection efficiency leading to electron leakage \cite{r8}. Auger recombination in the QW is however a somewhat favored mechanism \cite{r3}, with possibly an early onset induced by the reduction in active volume due to current crowding \cite{r9} or by carrier localization \cite{r10, r11}. Up to now, the evidence of this process comes from analysis of carrier dynamics \cite{r3}, either continuous wave or by time resolved measurements where the $n^2p$ or $np^2$ dependence of recombination rate is observed ($n$ and $p$ are the electron and hole concentrations respectively). Additional support for Auger comes from the fact that some possible cure to Auger seem to offer diminished droop phenomenon, thus enabling operation at higher current densities, such as using thicker active regions, i.e. thick QWs or even double heterostructures which reduces carrier densities \cite{r14}. However, the remedies do not yield unambiguous identification of the droop origin, as several mechanisms can be impacted by a given change in LED design - for instance reducing carrier concentration could also diminish carrier leakage \cite{r8}.

There is so far no direct evidence of the Auger carrier recombination mechanism in semiconductors by observing the remaining higher energy particle. Such direct observation would require a spectroscopic measurement of hot electron in the device. Spectroscopy of low energy electrons emitted into vacuum is a classical method to study hot electrons in semiconductors. Already in 1967, Eden \textit{et al.} \cite{r14a} measured hot electron emission spectra under excitation with visible light. Since that time, this technique has been widely used to study the electronic structure \cite{r18a, *r18b, *r18c} and hot electron transport properties \cite{r16a, r17} of various semiconductors and junctions. Multivalley transport toward the surface was predominantly observed due to efficient transfer to long lived side conduction valleys. This phenomenon was also recently evidenced in GaN \cite{r20} and AlN devices \cite{r14b}.

In this letter we report on the direct measurement of Auger electrons generated by carrier recombination in semiconductors by electron emission spectroscopy. The experiments were performed on GaN-based LEDs. Energy analysis of electrons emitted from the device into vacuum is performed as a function of forward bias current. The signature of Auger electrons is observed through high energy peaks which appear in the electron energy distribution curves (EDCs) at high injected current densities. The Auger electron current is found to correlate with the simultaneously observed droop in emission efficiency.

The principle of our experiment is shown in Fig.~\ref{fig:1}a. Electrons and holes are injected in the active layers (InGaN QWs) of a LED. The p-GaN surface is activated by cesium deposition to negative electron affinity (NEA) where the minimum of the conduction band (CB) in bulk p-GaN lies above the vacuum level at the surface. Electrons reaching the surface can be emitted into vacuum where their energy distribution is measured. During the transport, different processes give rise to different contributions in the emitted electron spectrum.

The origin of thermalized electrons emission can be several-fold. Cold electrons can be injected in the p-side at low energy by either bypassing capture into the QWs or by overflowing the QWs and subsequently overcoming the electron barrier layer (EBL) either by tunneling or by thermionic emission (process II in Fig.~\ref{fig:1}a). These `overshoot' electrons that reach the surface (a significant fraction will be lost by recombination in the 200~nm thick p-region) will largely be thermalized, even those electrons launched in the p side over the EBL with an energy in the fraction of eV range.  Indeed, the LO phonon emission time is 9~fs in GaN \cite{r15} which and is much shorter than the $\sim$1~ps transit time for electrons with thermal velocity of a few 10$^5$~m/s. Photoemission excited by the LED light, below the GaN bandgap \cite{r16}, reabsorbed near the surface (process IV in Fig.\ref{fig:1}1a) either due to impurity band transitions or to Franz-Keldysh transitions, could also generate low energy electron emission.

Highly energetic electrons can be created by Auger recombination in the QWs through an electron-electron-hole process which launches electrons with an initial energy equal to the recombining e-h pair in the QW, typically 2.7~eV in the blue (process I in Fig.~\ref{fig:1}a). The Auger process can be direct, or mediated through a phonon or by disorder (Fig.~\ref{fig:1}b). Theoretical calculations are still controversial \cite{r12, r13}.  As a result, Auger electrons may populate different valleys of the conduction band (process III in Fig.~\ref{fig:1}a). If such electrons do not fully thermalize in the $\Gamma$ conduction band before reaching the p-GaN surface, observation of hot electrons is expected in the EDC.

In the present study the sample was a GaN-based LED structure from Walsin Lihwa (Taiwan), grown by metalorganic chemical vapor deposition on a flat (0001) sapphire substrate. It consists of a several $\mu$m thick undoped buffer layer, followed by a Si-doped n-type buffer, an 8 period In$_{0.18}$Ga$_{0.82}$N/GaN multiple QW (3~nm thick InGaN QWs and 10~nm thick GaN barriers,) a 40~nm thick Al$_{0.15}$Ga$_{0.85}$N EBL, and a top 200~nm Mg-doped p-layer ([Mg] approximately 1.8$\times$10$^{20}$ cm$^{-3}$). The n-side terminal is a Ti/Pt electrode. The sample has a p-side square Pt electrode (side 500~$\mu$m) with an array of 27$\times$27 holes of 10~$\mu$m diameter each to expose the p-GaN.

The p-GaN surface was prepared in NEA with moderate temperature treatment. The sample was first treated in an HCl-isopropanol solution \cite{r16} and then introduced in a UHV setup designed for low energy electron spectroscopy of electron emission from semiconductors \cite{r17}. After annealing at 260$^\circ$C for  $\sim$30~min, the p-GaN surface was cesiated. The surface activation by cesium deposition was optimized by monitoring the electron emission current at injection current density of 0.4~A/cm$^2$. NEA was achieved without oxygen exposure.  The work function $\phi_{GaN}$ was 2.3~eV and remained stable for several days. Electrons emitted from the junction were energy analyzed (Fig. \ref{fig:1}c) with a resolution of 50~meV in a 90$^\circ$ electrostatic cylindrical deflection selector \cite{r18a, *r18b, *r18c}. The spectrometer was set in the constant path energy mode and the spectrum was obtained by scanning the sample potential (p-contact potential) V$_{cath}$. In this operation mode, the selected electrons are those that may enter with zero kinetic energy in a grounded gold surface (Faraday cup) of work function $\phi_{Au}\sim$4.8~eV. Their kinetic energy at emission is then: E$_k$ = $\phi_{Au}$ - $\phi_{GaN}$ - eV$_{cath}$. The device was biased by applying a potential V$_{bias}$ + V$_{cath}$ to the n-GaN contact. For high current density measurements, pulsed current injection was used with a 5\% duty cycle. We have checked that the collected current was proportional to the duty cycle.

The overall collection and transmission efficiency of the spectrometer and electron optics is  $\sim$10$^{-3}$. All EDCs were corrected by multiplication by the ratio of the total emitted current to the integrated measured current at the Au Faraday cup. We have normalized the energy level of the LED structure at the QW position by subtracting the ohmic drop in the n and p regions from the applied bias voltage as determined from the I-V characteristics. 

\begin{figure}[tbh]
\includegraphics[width=8cm]{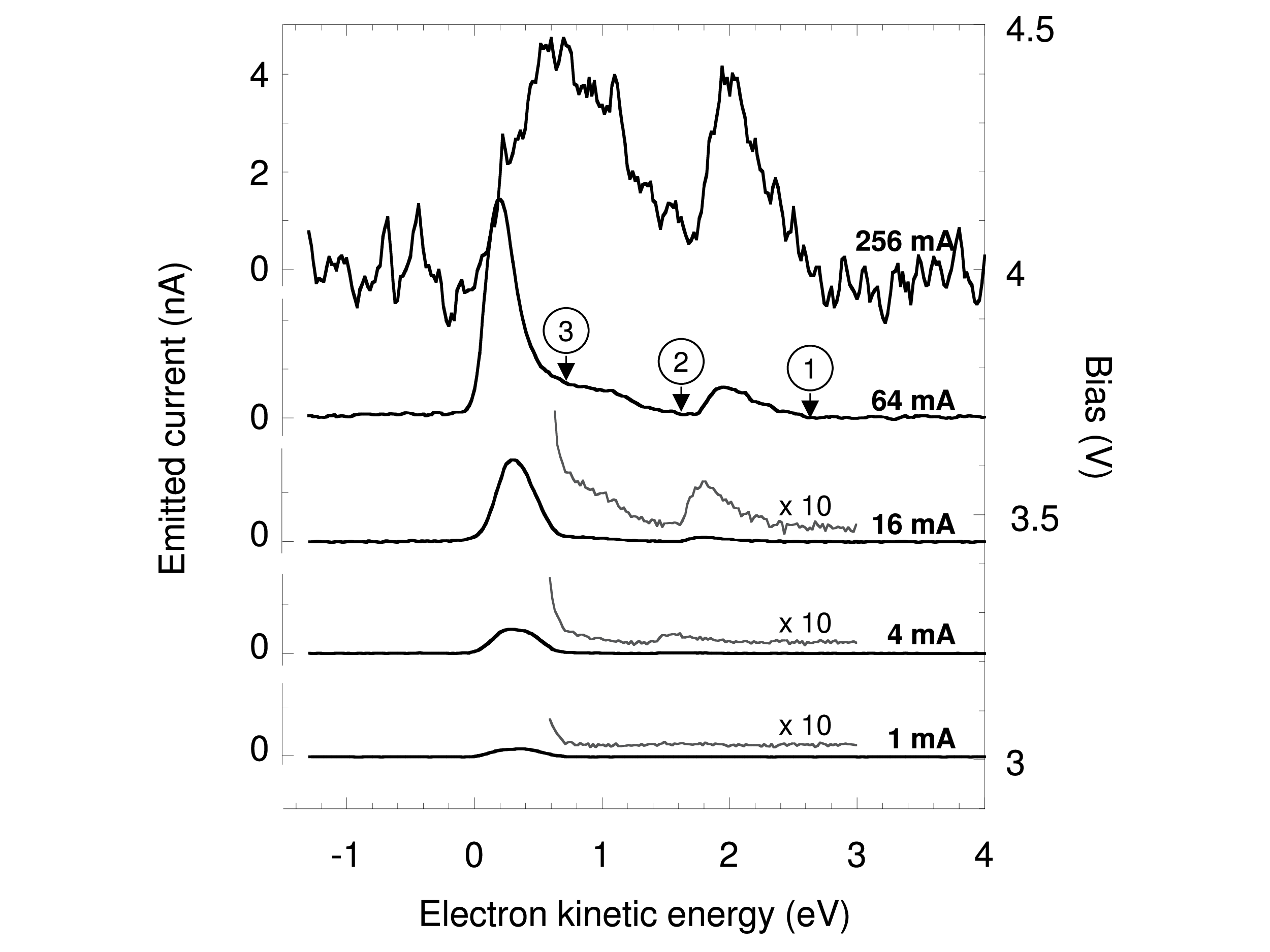}
\caption{\label{fig:2} Energy Distribution Curves (EDCs) for different injection currents. The base line of each spectrum was shifted by an amount proportional to the LED bias potential for that current (right-hand scale). Electron energy is referred to the vacuum level which does not change with bias and determines the low energy onset. When increasing injected current, high energy peaks appear around 2~eV signaling generation of hot carriers in the structure.}
\end{figure}

Figure~\ref{fig:2} shows the electroemission spectra measured at room temperature for different injected current. The noisy appearance of electron emission peaks at high injected current has two origins. First, the pulsed current injection with a 5\% duty cycle reduces the time averaged emitted current by a factor of 20. Second, blurring of the emitted electron beam occurs, due to the stray electric fields from the n and p electrodes, which strongly diminishes the collection efficiency in the electron optics. This results in a reduction of the effective collected current intensity. The EDCs normalization procedure described above does correct this signal reduction but it cannot increase signal-to-noise ratio. 

The electron energy is referred to the vacuum level at the p-GaN emitting surface which is 2.3~eV above the Fermi level. As usual, if we assume that the surface band bending region (BBR) amplitude is large enough so that the CB minimum at the p-GaN surface is below the vacuum level, the low energy onset of the emitted electron spectrum lies at the vacuum level position \cite{r18a, *r18b, *r18c}. 

At current below 1~mA (current density below 0.4~A/cm$^2$), a single low energy emission peak was observed and corresponds to thermalized electrons, either injected into the p-side of the device junction, which were thermalized in the CB and subsequently underwent some further thermalization in the BBR or to photoemission excited by the LED light in the BBR. Because of the presence of an electron blocking layer, it is more probable that this low energy peak is due to LED light excited photoemission from the BBR. This interpretation is supported by the observation of a similar photoemission peak (not shown here) under external excitation with a laser light of energy close to the LED emission.

Higher energy peaks appeared at 4~mA injected current and higher.  The main peak was $\sim$1.5~eV above the low energy peak. A somewhat weaker intermediate peak was also observed at 0.3-0.4~eV above the low energy peak.  The relative intensities of these two high energy peaks remain in the same ratio with increasing current, thus proving their common origin. The high energy threshold of the highest energy peak lies about 1.1~eV above the minimum of the CB in the bulk p-GaN region \cite{r19}.

\textit{The only viable mechanism to generate high energy carriers in the structure is Auger recombination}.
The bias potential dropped in the LED junction is close to the flat band potential and cannot produce hot electron injection at high energies. The other ways to generate hot carriers would be hot carrier launching by an energy barrier in the structure or carrier heating by high electric fields. The former mechanism requires a barrier with an energy discontinuity that does not exist in the LED structure. The latter mechanism requires a region in the biased LED with a very high electric field but no such region exists in the LED, with the exception of the BBR. However, acceleration in the BBR electric field cannot promote electrons at higher total energy. Indeed, although electrons may gain kinetic energy in the BBR, their total energy at the surface obviously remains smaller than in the bulk.

The available data on hot carriers in GaN supports this interpretation of the high energy peak as due to another CB valley (for simplicity called here $L$). Transport measurements point to the existence of such a band, however with quite some scatter about its energy position (see e.g. \cite{r21}). Given the fast LO phonon emission in GaN, it is not possible to directly observe Auger electrons at their initial kinetic energy after traversing 200~nm of p-GaN. Actually, the phonon-assisted Auger process may directly yield Auger electrons within the $L$-band (Fig.~\ref{fig:1}b right). From their initial high energy in that band, electrons would thermalize quickly to the bottom of the $L$ band:  we expect phonon relaxation to be extremely fast in that band due to the high density of states and the many valley states. In InP, for instance, intra-side-valley phonon scattering is $\sim$4 times faster than phonon emission in the $\Gamma$ valley \cite{r17}. Hence, we consider that the bottom of the $L$ valley acts as a source of thermalized electrons for emission into the vacuum, as was observed in GaAs \cite{r14a}, InP \cite{r18a, *r18b, *r18c} Si \cite{r16a} and AlN \cite{r14b}. This source emits an electron peak which high energy onset lies at the bottom of the $L$ valley band in the bulk p-GaN (kT is not observable in our setup). As the $L$ valley minimum follows the BBR potential, electrons moving towards the surface relax part of their energy and the peak broadens towards lower energies by the energy of the few phonons emitted when traversing the BBR.  If the Auger process would occur in the $\Gamma$ valley (Fig.~\ref{fig:1}b), electrons would also transfer in the $L$ valley before reaching the surface as the time transfer from $\Gamma$ to $L$ valley is very fast, at most 170 fs \cite{r22} as observed on the transfer threshold from $\Gamma$ valley, but even faster from higher energy electrons in the $\Gamma$ band. The experiment so far does not distinguish between the two possible Auger mechanisms.

The intermediate band, 300~meV above the lower peak, connected to the $L$ valley emission, can originate from electrons scattered out of the $L$ valley into the $\Gamma$ band near the surface and thermalized at the bottom of the p-GaN conduction band.

\begin{figure}[tbh]
\includegraphics[width=8cm]{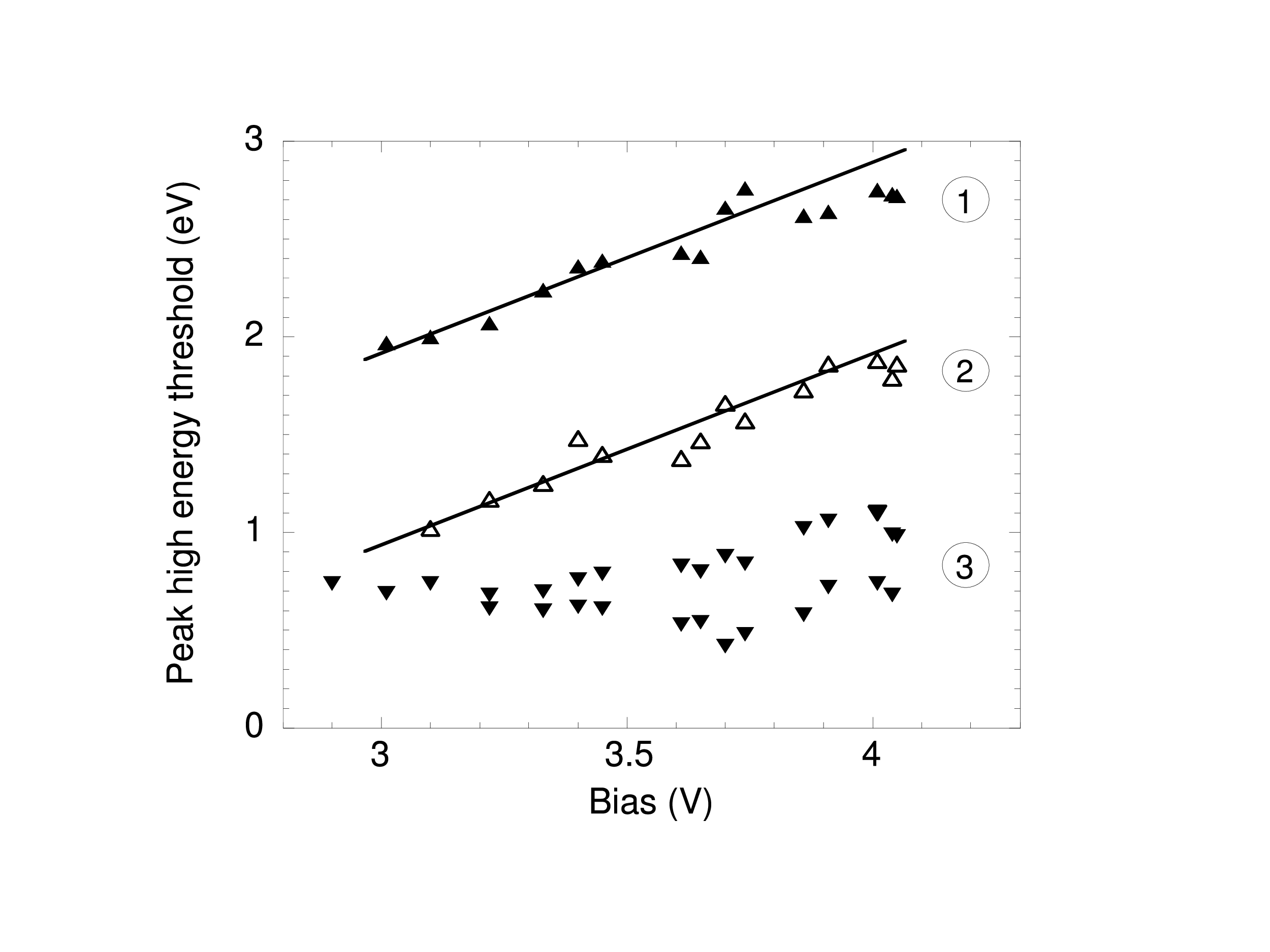}
\caption{\label{fig:3} Energy position of the three peak thresholds (see Fig.~\ref{fig:2}, points 1,2,3) under changing bias. As shown by the straight lines with slope 1, the hot electron energy increases with the applied bias. When the the flat band potential in the p-n junction has been reached, increasing injected current requires a voltage drop in the BBR near the p-GaN surface, resulting in an increase in the position of the bulk energy levels with respect to the surface ones.}
\end{figure}

Analyzing the details of the results, the high energy features shift in energy with changing bias (Fig.~\ref{fig:3}). This shift is due to the rectifying character of the p-GaN contact which drops most of the bias once flat band potential is reached in the p-n junction \cite{r25}. Thus, the positions of bulk p-GaN bands (in particular of the $L$ valley) relative to the constant p-contact potential (i.e. to the vacuum level) increases when increasing bias from 3~V to 4~V.

Electron emission in vacuum from forward biased GaN p-n junctions was previously observed by Shaw \textit{et al.} \cite{r26} with currents up to 5~A/cm$^2$. As expected, no new high energy peak was observed there as the carrier density, distributed over the carrier diffusion length, typically 100~nm, is quite smaller than in the 3~nm thick InGaN QWs of our sample.

\begin{figure}[tbh]
\includegraphics[width=8cm]{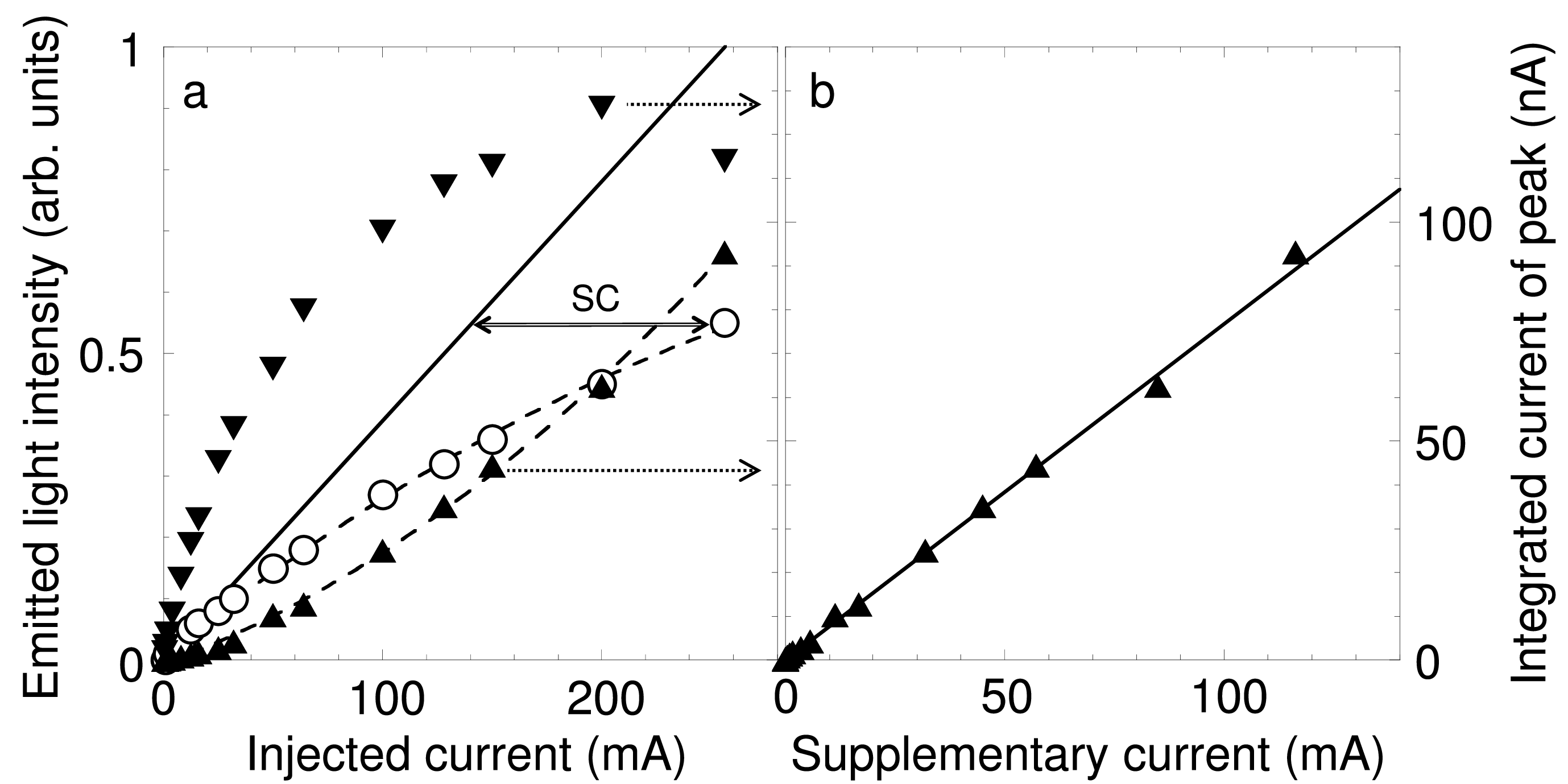}
\caption{\label{fig:4} (a) Plots of the integrated low  ($\blacktriangledown$)and high ($\blacktriangle$)  electron energy peaks current, and of the optical power ($\circ$) out as a function of injected current. The straight line is the expected optical output power in the absence of droop, obtained by a linear extrapolation from low currents.  (b) Plot of the integrated current over the high electron energy peak as a function of the supplementary current (SC), i.e. the diﬀerence between the actual injected current and the current which would give the same optical output if the low current light emission efficiency had been conserved.}
\end{figure}

Simultaneously along with the measurement of the electroemission current and spectrum, we have measured the light intensity emitted by the LED (Fig.~\ref{fig:4}a). The ``Auger electron current'' (integrated high energy peak current) correlates with the ``droop current'' component as deduced from the dependence of optical power on input current (Fig.~\ref{fig:4}b). We consider the ``droop current'' as the ``supplementary current'' (labeled SC in Fig.~\ref{fig:4}a) necessary to obtain a given light output power (empty circles in Fig.~\ref{fig:4}a) when compared to that expected from a linear extrapolation (full line in Fig.~\ref{fig:4}a) from low current low output power to high currents. 

The electron emission experiments presented here strongly supports the observation of Auger electrons. However, a question remains: is it the dominant droop mechanism? There are two pieces of evidence that convince us that it is the case. First, if another mechanism were responsible for the droop, its effect would set in at lower current than Auger generation, so that Auger is only a minor cause of the disappearing electrons at the onset of droop. The concurrent appearance of Auger electron emission and the onset of droop shows that Auger is indeed the major cause for droop. Second, the electron current emitted in vacuum has the right order of magnitude to account for the disappearing recombination current: we measure (Fig.~\ref{fig:4}b) a total electron emission current of 80~nA for a supplementary current in the LED of 100 mA, thus an efficiency of $\sim$10$^{-6}$. We can evaluate an efficiency in that range. First, most of the current is injected below the p-electrode, and therefore does not yield any outside current due to the emission masking by the electrode. Only the fraction of injected current within a current spreading length, on the same order as the thickness of the p-GaN, from the edge of the unmasked apertures contributes to electron emission. For a 200~nm current spreading length, only 1\% at most of injected current participates in emission. Second, only a fraction of the unmasked Auger electrons can be observed: half of them are emitted in the direction opposite to the surface; some of the Auger electrons transferred to the $L$ valley do not reach the surface as they undergo reverse transfer to the $\Gamma$ valley (scattering time from $L$ to $\Gamma$ is 1~ps \cite{r22}), and a significant fraction of such $\Gamma$ electrons recombine in the 200~nm p-layer. The GaN emission quantum yield is low as the cesiation was not optimized and the cesiated surfaces were used for a number of days, leading to a quantum yield in the $\sim$10$^{-3}$ range. Another cause for a reduction in observed current is the recapture of electrons by the p electrode before they are attracted by the spectrometer entrance slit.

In conclusion, we have directly observed for the first time, the generation of Auger electrons under electrical carrier injection in a semiconductor by the energy analysis of electrons emitted in vacuum from a p-n junction. In the studied structure, an InGaN LED, the measurement unambiguously assigns the droop in quantum efficiency observed at high injection current densities to Auger recombination of carriers and therefore brings essential information in a long standing controversy. 

\begin{acknowledgments}
The samples were kindly provided by H.T. Chen, K.M. Chen, and S.-C. Huang from Walsin Lihwa (Taiwan). The authors thank Y. Lassailly for fruitful discussions. The work at UCSB is supported by the Center for Energy Efficient Materials (CEEM), an Energy Frontier Research Center funded by the U. S. Department of Energy, Office of Science, Office of Basic Energy Sciences under Award Number DE-SC0001009.
\end{acknowledgments}

%


\begin{thebibliography}{29}
\bibitem [{r25()}]{r25}%
  \BibitemOpen
  \href@noop {} {}\bibinfo {note} {Ohmic potential drop in the p-GaN region
  would not produce an energy shift of the Auger peak (electrons in the $L$
  valley will follow the progressive potential drop across the p-GaN
  thickness). Ohmic potential drop at the p-contact is already taken into
  account and produces a shift of the low energy edge. The only possibility for
  shifting the Auger peak is thus a rectification at the metal / p-GaN
  contact.}\BibitemShut {Stop}%
\end{thebibliography}
\end{document}